\documentclass[%
 reprint,
 amsmath,amssymb,
 aps,
pra,
]{revtex4-2}

\usepackage{graphicx}
\usepackage{dcolumn}
\usepackage{bm}

\usepackage[utf8]{inputenc}
\usepackage[T1]{fontenc}
\usepackage{mathptmx} 

\usepackage{graphicx}
\usepackage{dcolumn}
\usepackage{bm}

\usepackage{verbatim}
\usepackage{hyperref}
\hypersetup{
    colorlinks=true,
    linkcolor=blue,
    citecolor=blue,
    urlcolor=blue,
    filecolor=blue
}
\usepackage{soul}
\sethlcolor{yellow} 
\usepackage{float}
\usepackage{orcidlink}
\begin{document}

\preprint{APS/123-QED}

\title{ Charge exchange reactions with central and tensor interactions and nuclear matrix elements associated with  $\beta\beta$-neutrinos and  astro-neutrinos}

\author{Hiroyasu Ejiri\,\orcidlink{0000-0003-0568-3528}}
\email{ejiri@rcnp.osaka-u.ac.jp}
\affiliation{Research Center for Nuclear Physics, The University of Osaka 565-0871, Japan
}%
\author{Yoritaka Iwata\,\orcidlink{0000-0002-6898-7505}}

 \affiliation{Osaka University of Economics and Law, Osaka 581-0853, Japan}



\date{\today}

\begin{abstract}
Nuclear matrix elements (NMEs) for neutrinoless $\beta\beta$ decays (DBD) and inverse-$\beta$ decays (IBD) are crucial for studying neutrino ($\nu$) properties beyond  and within the standard model by using DBDs and  IBDs.  The charge exchange nuclear reaction (CER) is a realistic probe for studying experimentally these NMEs. The present work shows for the first time an experimental way to extract the Gamow-Teller (GT) and tensor (T) NMEs associated with DBDs and IBDs from the $q$ (momentum transfer) dependence of the CER cross-section by taking into accounts explicitly both the central and tensor interactions of $V^{\rm C}$ and $V^{\rm T}$. The present method was applied for CERs on $^{71}$Ga used for the solar-$\nu$s and the Ga-$\nu$ anomaly and on $^{100}$Mo and $^{128}$Te used for DBD and IBD $\nu$s. Impacts of the CER with the tensor interaction on the DBD and IBD NME studies are discussed.
\end{abstract}


\maketitle


{\it Nuclear matrix elements for neutrino studies.} 
Neutrino ($\nu$) properties such as the Majorana $\nu$-mass, the right-handed weak interaction and others beyond the standard electro-weak model (SM)  are studied by measuring  neutrinoless double beta decays (DBDs)\cite{eji05,avi08,ver12,ago23}. 
Astro-$\nu$ productions,  syntheses and  oscillations are investigated by using astro-$\nu$ nuclear reactions, i.e. inverse-$\beta$ decays (IBDs) \cite{eji00,eji02,eji19}. 

DBD and IBD amplitudes (square roots of their decay rates) are proportional to the nuclear matrix elements (NMEs) for DBDs and IBDs. The present letter addresses the NMEs, which are crucial for DBD-$\nu$ and astro-$\nu$ studies beyond and within SM
 \cite{
gys19,suh98a,fae98,men11,suh12,bar13,suh17,eng17,jok18,eji22,eji23,eji25}.  The NMEs for $^{71}$Ga are of great interest in view of the Ga anomaly and the possible sterile $\nu$ \cite{abd06,kae10,giu11,bar22a,bar22b,ell23,ell24,bas21,giu21,hax98,fre11,kos19,iwa26}.

Actually,  medium heavy nuclei used for DBDs and IBDs are the complex many-body system composed by $A\approx 100$ baryons with $A$ being the mass number.  Thus the DBD and IBD NMEs are very sensitive to all kinds of nucleonic and non-nucleonic correlations among the many baryons (nucleons, isobars) in the DBD and IBD nuclei and thus evaluated NMEs scatter over a factor around 5-10, depending on the models and the parameters used there \cite{eji00,eji19,gys19,suh98a,fae98,men11,suh12,bar13,suh17,eng17,jok18,eji22,eji23,eji25} . Recently the $\Delta$ baryon is shown to quench the axial-vector NMEs \cite{eji25}. Note that the excited states in wide ranges of the excitation energy and the multipolarity are involved in the DBD and  IBD NMEs. Then, accurate evaluations of the DBD and IBD NMEs by using only theoretical nuclear models are very hard. 

Accordingly, experimental studies of the NMEs are crucial for evaluating accurately the DBD and IBD NMEs and/or to help nuclear theories to calculate accurately them.   "Electro-weak"  $\nu$  e $\gamma$ probes may be in principle used to study the present weak interaction NMEs, but are not in practice.   $\nu$-beam experiments are not realistic  because of the extremely small cross-section. Photo-nuclear and e-beam excitations include both isovector and isoscale NMEs, and can not access the present isovector DBD and IBD NMEs. NMEs studied by EC/$\beta$ are limited to the ground state in special EC/$\beta$ nuclei. No reliable GT NMEs for DBD nuclei of current interests
are available from EC/$\beta$.
\begin{figure}[ht]
\begin{center} 
\hspace{+2.3cm}
\vspace{-.4cm}
\includegraphics[width=0.5\textwidth]{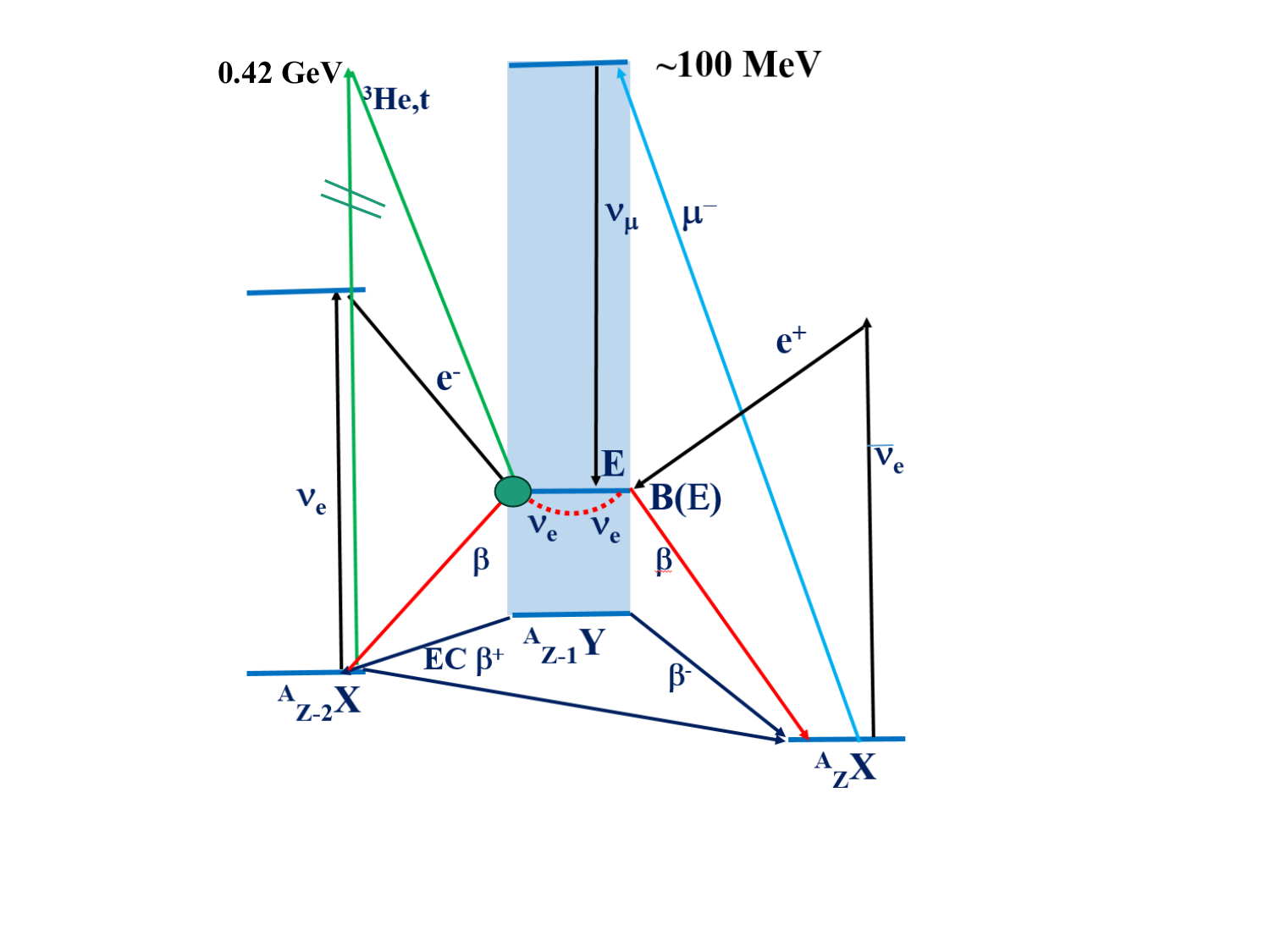}
\vspace{-1.6cm}
\end{center} 
\caption{DBD and IBD schemes and RCNP CER and OMC \cite{eji19}.  The green circle is the NME to be studied by the present CER. 
\label{figure:fig1}} 
\vspace{-0.2cm}
\end{figure}

The present letter aims to show  that  nuclear charge-exchange reactions (CERs) of $^A_Z$X$\rightarrow^A_{Z\pm1}$Y are well used for  studying experimentally the NMEs for DBDs of $^A_Z$X$\rightarrow^A_{Z\pm2}$Z via the intermediate nucleus $^A_{Z\pm1}$Y and those for IBDs of  $^A_Z$X$\rightarrow^A_{Z\pm1}$Y \cite{eji00,eji19}.  Actually,
the ($^3$He,$t$) CER with the intermediate-energy projectile at Research Center for Nuclear Physics (RCNP) Osaka has been extensively used for studying  the $\tau^-$-side DBD and IBD NMEs for the ground and excited states up to 30 MeV \cite{eji00,eji19,fre11,iwa26,aki97,thi12,thi12b,pup11,gue11,thi12a,pup12,fre13,eji16a,dou20}.
The GT NME of $M$(GT)=$<\tau\sigma>$ has been derived from the CER cross-section of $d\sigma/d\Omega=k|V^CM(\rm GT)|^2$ with $V^{\rm C}$ being the central $\tau\sigma$ nuclear interaction \cite{eji00,eji19,goo80,tad81,goo84}. The corresponding weak NME is g$_{\rm A}M(\rm GT)$ with g$_{\rm A}$ being the axial-vector weak coupling and $M$(GT) being the same nuclear $\tau\sigma$ NME of the CER.

One critical issue of the "nuclear" CER, is the tensor term $V^{\rm T}M$(T) with $V^{\rm T}$ being the tensor interaction and $M$(T)=<[Y$_2\sigma]_1$> being the tensor NME. Note that $V^{\rm T}$ is  due to the Meson exchange current, being characteristic of the nuclear interaction.
  Although the $V^{\rm T}$ is much smaller than the $V^{\rm C}$,  the interference effect gets appreciable unless $M$(T)$\ll M$(GT).  Then the CER without the tensor interaction effect is used only for special GT states with very large $M$(GT). The contribution of the tensor interaction for the non-GT(i.e. l-forbidden) transition has been discussed theoretically \cite{hax98,kos19,iwa26}. 

The present letter shows for the first time an experimental way to extract NMEs relevant to DBDs and IBDs from CER experiments by properly correcting for the tensor interaction effect.
We note that  $V^{\rm C}$ decreases slowly 
with increase of the momentum transfer $q$, while the $V^{\rm T}$ due mainly to the pion exchange increases steeply with increase of $q$, getting maximum at around $q\approx$140 MeV/c. Thus  the effect of the $V^{\rm T}M$(T) is studied by measuring the $q$ dependence of the CER cross section. The $q$ dependence shows up in the angular distribution of the CER cross section on the scattering angle $\theta$ since the $q$ increases almost linearly as the angle $\theta$ increases. 

 We first discuss the $q$ dependences of $V^{\rm C}$, $V^{\rm T}$, and  the 
 CER  cross-sections, and then a realistic way to extract the $M$(GT)  relevant to the DBDs and IBDs. Then  we discuss  the CER on $^{71}$Ga used for the solar $\nu$ studies and the Ga anomaly and the CREs on the IBD and DBD nuclei of  $^{100}$Mo and $^{128}$Te.\\

{\it The central and tensor interactions in the CER}. ~The nuclear interaction is expressed as 
$V^{\rm eff}= V^{\rm C}+V^{\rm T}$. The spin orbit (LS) interaction at the present region of $q\ll$200 MeV/c  is so small that we do not consider it here. The central and tensor interactions are given by \cite{bro81,fra85,fay97}
\begin{eqnarray}
V^{\rm C}=V^{\rm C}(r_{ij})+V^{\rm C}_{\sigma}(r_{ij})\sigma_i\sigma_j+~~~~~~~~\\
V^{\rm C}_{\tau}(r_{ij})\tau_i\tau_j+V^{\rm C}_{\sigma\tau}(r_{ij})\sigma_i\tau_i\sigma_j\tau_j, \nonumber \\
V^{\rm T}=[V^{\rm T}(r_{ij})+V^{\rm T}_{\tau}(r_{ij})\tau_i\tau_j]S^T_{ij},~~~~~~~~
\end{eqnarray}
where $S^T_{ij}$ is the tensor operator. 
The $V^{\rm C}$ and $V^{\rm T}$ at the incident energy of $E_{\rm p}$=140 MeV (i.e. $E(^3$He)=0.42 GeV) are plotted as a function of $q$ in Fig.1. The  $V^{\rm C}$ decreases slowly  as $q$ increases, while $V^{\rm T}$ increases rapidly as $q$ increases. Then the ratio of $V^{\rm T}$/$V^{\rm C}$ increases rapidly as $q$ (see Fig.1). 
\begin{figure}[ht]
\begin{center} 
\hspace{-0.3cm}
\vspace{-0.4cm}
\includegraphics[width=0.35\textwidth]{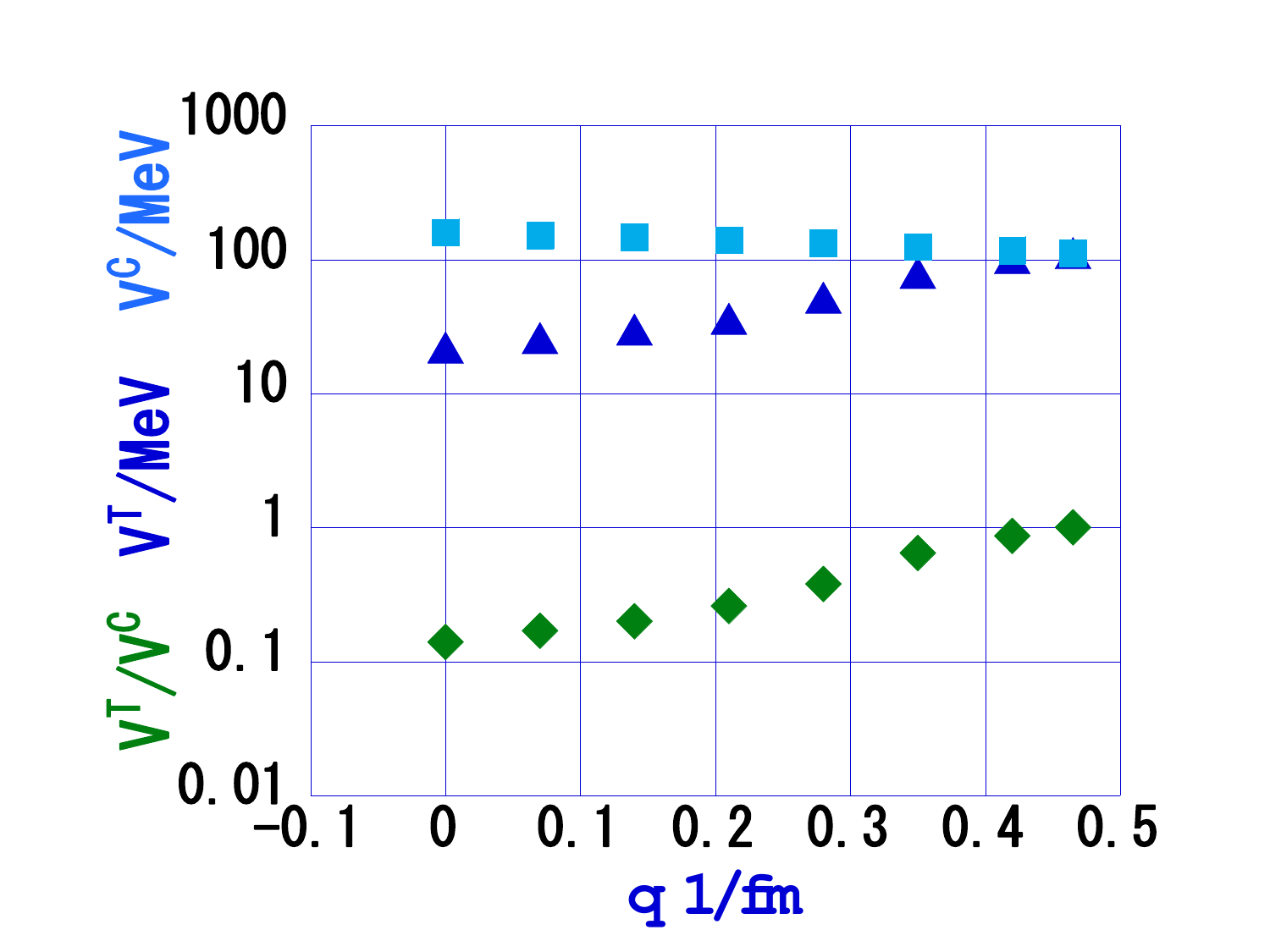}
\vspace{-0.3cm}
\end{center} 
\caption{The central and tenser interactions of $V^{\rm C}$ (light blue squares) and $V^{\rm T}$ (blue triangles) in units of MeV  and the ratio of $V^{\rm T}$/$V^{\rm C}$ (green diamonds) against  $q$ in units of fm$^{-1}$. 
\label{figure:fig1}} 
\vspace{-0.cm}
\end{figure}

The CER differential cross section is given by \cite{eji00,eji19}
\begin{equation}
\frac{d\sigma(q)}{d\Omega}=\Sigma_{i,j,k}\frac{d\sigma(ijk,q)}{d\Omega},
\end{equation}
where $i,j,k$ are the projectile, target and relative angular momentum transfers. The present work is concerned mainly with the GT ($i$=$j$=1, $k$=0) excitation. Thus the differential cross section to be considered is the ($i,j,k$)=(110) component with the s-wave ($\Delta l$=0) out-going $t$.  The angular distribution is given approximately by the square of the rank-0 spherical Bessel function of $j_0(qR)$ with $R$ being the effective nuclear radius.  It shows the forward-peaking $q$ distribution, getting maximum at $q$=0.  Hereafter we mainly discuss the differential cross section of d$\sigma (110,q)$/d$\Omega$=$D(110,q)$.

$D(110,q)$  does include both the $M$(GT) by the central interaction of $V^C_{\tau\sigma}$ and the $M$(T) by the tensor interaction of $V^T_{\tau}$.  Then $D(110,q)$ is expressed as 
\begin{eqnarray}
D(110,q)=K(k_ik_fE_iE_f) (j_0(qR))^2 N^{\rm D} |J^{\rm C}(q)|^2 B(110), 
\\ 
B(110)=B({\rm GT})(1+f(q))^2,~~~~~~~~~~~~~~\\
f(q)=\epsilon \frac{M(\rm T)}{M(\rm C)}\frac{V^{\rm T}(q)}{V^{\rm C}(q)} , ~~~~~~~~~~~~~~~~~~~
\end{eqnarray}
where $K(k_ik_fE_iE_f)$ is the kinematical factor, $N^{\rm D}$ is the distortion factor, $J^{\rm C}(q)$ is the interaction volume integral, $B(110)$ is the reduced width, $B({\rm GT})$ is the GT reduced width, and $f(q)$  is the correction factor for the tensor interaction contribution and $\epsilon$ is the tensor effect coefficient.  For simplicity, we set $V^{\rm C}_{\tau\sigma}=V^{\rm C}$ and $V^{\rm T}_{\tau}=V^{\rm T}$.
The $B$(GT)=(2$J_{\rm i}+1)^{-1}(M({\rm GT}))^2$ is the GT reduced width with $J_{\rm i}$ being the initial state spin.  So far, CER experiments have assumed pure GT excitations without the tensor contribution, namely $f(q)$=0 \cite{eji00,goo80,tad81,goo84}.

Then the contribution of the tensor NME is derived  experimentally from the $q$ distribution of the differential cross-section.  Since the CER $Q$-values are nearly 0 for low lying states with the excitation energy $Ex\ll$0.42 GeV, $q$ increases linearly as the scattering angle $\theta$ increases. 

The differential cross section for the $i$ th final state is given by rewriting the eqs.(4-6) as 
\begin{eqnarray}
D_i(110q)=K'd_i(110q) B_i({\rm GT}), ~~~~~
\\ ~~
d_i(110q)=(j_0(qR))^2(1+f_i(q))^2,~~~~~
\end{eqnarray}
where $K'=K(k_ik_fE_iE_f)  N^{\rm D} |J^{\rm C}(q)|^2$ and $d_i(110,q)$ is the $q$ distribution. Here we assume the common $K'$ for all low-lying states of $i$ since their excitation energies $E_{\rm X}$ are much smaller than the incident and scattered particle energies. The $q$ distribution of $d(110q)$ deviates more or less from the kinematic  distribution, which is approximately given by of $(j_0(qR))^2$ in case of the small distortion, by the factor F=$(1+f_i(q))^2$,  as shown in Fig.3. 

The effect of the tensor contribution is seen experimentally in the $q$ dependence of the differential cross section as given in Eqs. (4-8). The $q$ dependence is given  by the two factors, one is the kinematical factor given by $|j_0(qR)|^2$ in case of the small distortion of the incident and outgoing particle, and the other is the tensor-interaction factor $(1+f(q))^2$. The relative cross sections with and without the tensor contribution of $f(q)$ in case of $\epsilon M^T/M^C$=$\pm$0.5
are shown in Fig. 4 

\begin{figure}[ht]
\begin{center} 
\hspace{0cm}
\vspace{-0cm}
\includegraphics[width=0.5\textwidth]{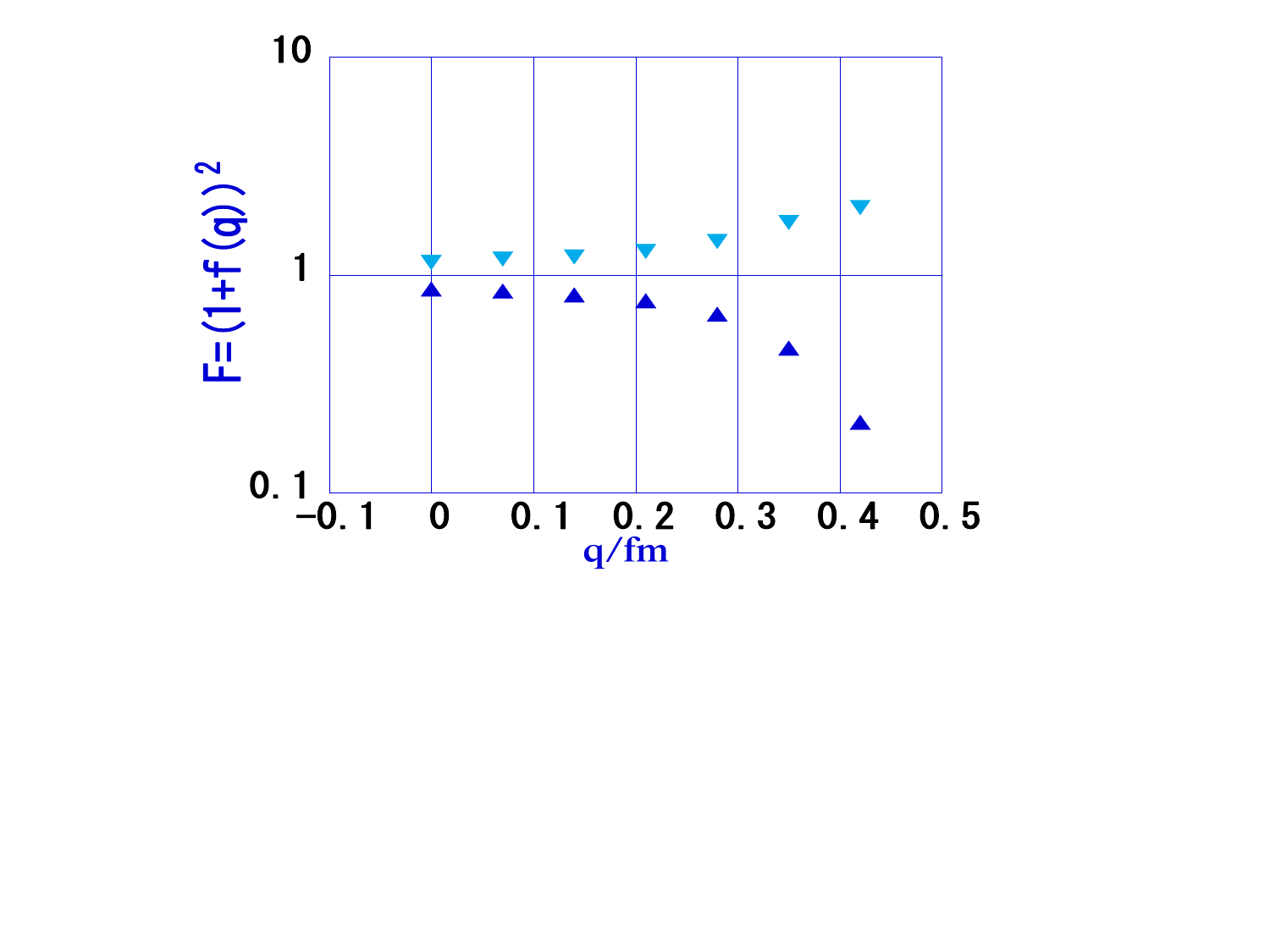}
\vspace{-3.6cm}
\end{center} 
\caption{The  tensor-interaction factor F=$(1+f(q))^2$ against  $q$.  Inverse triangles for $\epsilon M^T/M^C$=+0.5. Triangles for $\epsilon M^T/M^C$=-0.5.
\label{figure:fig1}} 
\vspace{-0.2cm}
\end{figure}

\begin{figure}[ht]
\begin{center} 
\hspace{0.4cm}
\vspace{-0cm}
\includegraphics[width=0.5\textwidth]{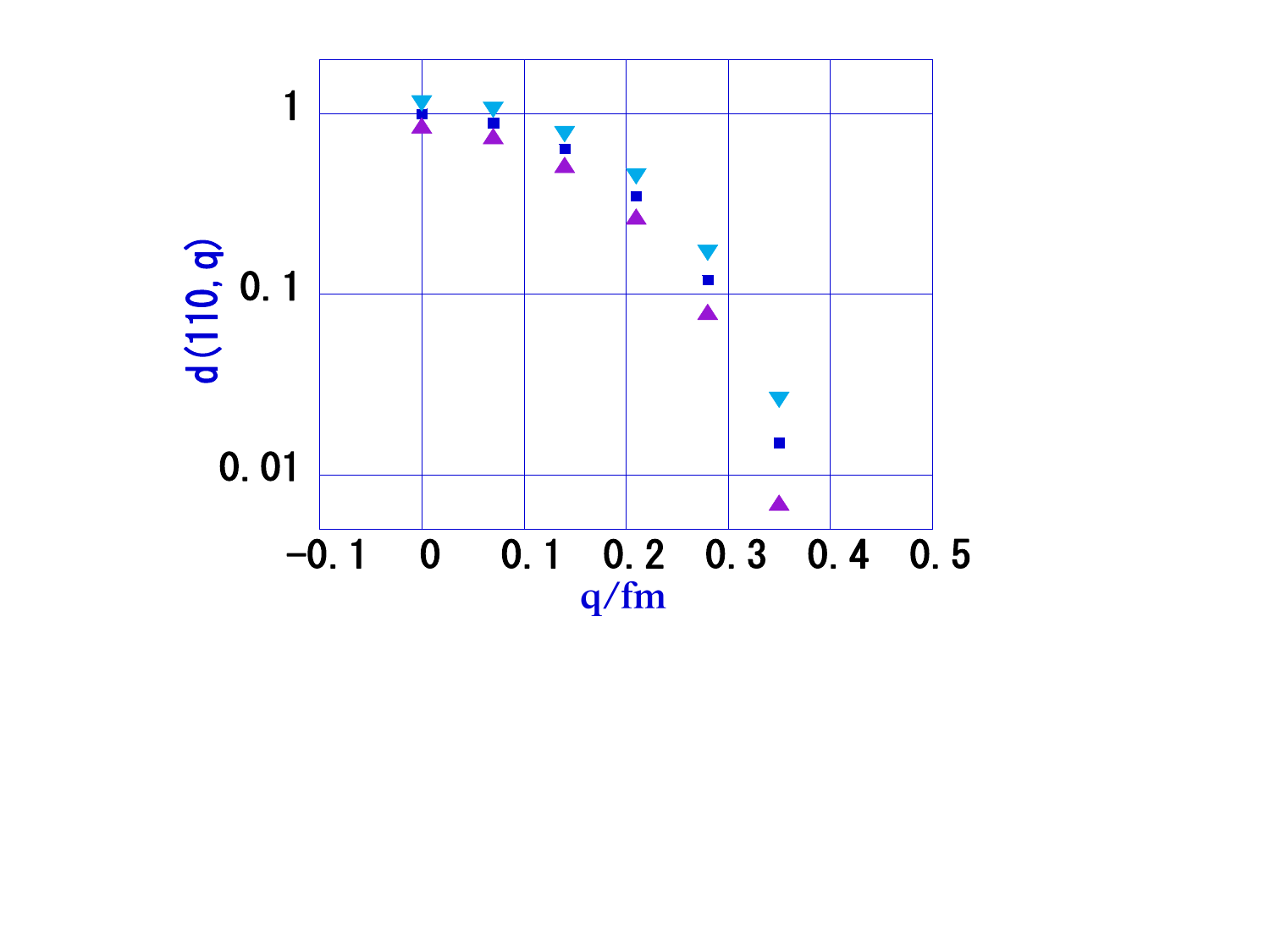}
\vspace{-3.3cm}
\end{center} 
\caption{The $q$ distributions of $d(110,q)$ against $q$ in unis of fm$^{-1}$. Light-blue inverse triangles, blue squares and violet triangles are the distributions with $\epsilon M^T/M^C$=+0.5, 0 and -0.5, respectively. 
\label{figure:fig1}} 
\vspace{-0.5cm}
\end{figure}

So far, the $B$(GT) value has been derived from the cross section $D(110,0)$ at $\theta$=0 ($q\approx$0) where $j_0(qR)$=1 by assuming no tensor contribution, namely $f_i(q)$=0 \cite{eji00,eji19}. However, the differential cross section of $D(110,0)$ should be corrected for the tensor contribution factor of $(1+f_i(0))^2$ to get $B$(GT) if $f(q)$ is appreciable even at $q\approx$0. We note that the correction factor of $f_i(q)$ is derived experimentally from the deviation of the cross section from the square of the Bessel function of $j_0(qR)$ as given in Eq. (8).



Actually, differential CER cross sections are conventionally presented not as the function of $q$, but as the function of the scattering angle $\theta$ since $\theta \approx q/k_{\rm i}$ in most realistic cases of $k_i \approx k_f$ with $k_i$ and $k_f$ being the incident particle momentum and the outgoing particle  momentum. \\

Now we discuss one realistic way to extract the GT and T  NMEs of $M$(GT) and $M$(T) in actual DBD and IBD nuclei by using the measured CER $q$ distributions. Actually, there are three kinds of i$\rightarrow$f transitions in medium heavy nuclei, (s):  spin-stretched quasi-particle (SQP) transitions of $l\pm1/2\leftrightarrow l\mp1/2$, (l): $l$-forbidden transitions of $l-1/2\leftrightarrow (l-2)+1/2$ and (m): other mixed transitions. In cases of (s) and (l), $M$(GT)$\gg M$(T) and $M$(GT)$\ll M$(T), respectively, while in cases of (m), $M$(GT)$\approx M$(T). Then we select one SQP GT transition (r) with a large $M_{\rm c}$(GT) as a reference GT NME, and discuss the NME for the $i$-th state by using  the ratio,  
\begin{eqnarray}
R_i(q)=\frac{\sqrt{D_i(110q)}}{\sqrt{D_{\rm r}(110q)}}~~~~~~~~
\\
=\frac{|M_i({\rm GT})|}{|M_{\rm r}({\rm GT})|}[|1+\epsilon \frac{M_i({\rm T})}{M_i({\rm GT})}\frac{V^{\rm T}(q)}{V^{\rm C}(q)}|],
\end{eqnarray}
where the tensor contribution $f(q)$ is assumed to be much smaller than the GT one for the reference transition (r), i.e. $\epsilon [(M_{\rm r}({\rm T}))/(M_{\rm r}({\rm GT}))][(V^{\rm T}(q))/(V^{\rm C}(q)])\ll1$. Then, noting that $V^{\rm T}(q)/V^{\rm C}(q)$ increases monotonically as $q$ increases, we plots the observed cross section ratio as a function of  $V^{\rm T}(q)/V^{\rm C}(q)$ in the forward angle region of $\theta$=0-2 deg.  Then the GT NME of $M_i({\rm GT})$ is derived from the ratio at $V^{\rm T}/V^{\rm C}$=0, and $\epsilon M_i({\rm T})/M_i({\rm GT})$ from the gradient.
Here the reference GT state is not necessarily a state in the same nucleus, but may be in a neighboring nucleus.\\





{\it GT and T NMEs by CERs on DBD and IBD nuclei}.
 NMEs relevant to the DBD and IBD neutrinos have been extensively studied  by using the CERs at RCNP as given in \cite{eji00,eji19}. 
 Let us apply the present analysis method for the CERs on $^{71}$Ga  used  for the solar-$\nu$ and the Ga-anomaly (sterile $\nu$) and  $^{100}$Mo used for DBD and IBD-$\nu$ studies \cite{eji19,eji02}. 
 
The CER of the $^{71}$Ga ground 3/2 state  to the $^{71}$Ge ground 1/2 state is mainly the spin stretched SQP transition of the 2p1/2$^-$ neutron$\rightarrow$2p3/2$^-$ proton \cite{iwa26}. The measured NME of $M$(GT)=0.585 is just the SQP one and agrees with the measured $\beta$-decay NME \cite{ele15}. Thus we take it as the reference cross-section  for the CERs on $^{71}$Ga. 
Then the CERs on $^{71}$Ga  to the 175 keV 5/2$^-$ and 500 keV 3/2$^-$ states in the $^{71}$Ge \cite{fre11,iwa26} were reanalyzed by explicitly taking into account the tensor contribution. The scattering angle there is converted to the ratio $V^{\rm T}$/$V^{\rm C}$ which increases as the scattering angle. Then the square root  of the cross section ratio (equ. 6) for the 175 keV and 500 keV states are given in Fig. 5. 

The square root of the cross section ratio increases  as the ratio $V^{\rm T}$/$V^{\rm C}$ (i.e. the momentum transfer $q$) increases for both the 175 keV and 500 keV states, clearly indicating the tensor contributions. The GT and T NMEs for the 175 and 500 keV states are derived  from the fits as given in Table 1. 

The CER of the $^{100}$Mo to the $^{100}$Tc 1$^+$ ground state is the spin stretched SQP transition of the 1g7/2$^+$ neutron $\rightarrow$1g9/2$^+$ proton. The measured NMEs of $M$(GT)=0.59  is just the SQP ones and agree with the EC/$\beta$ data. Thus we take it as a reference cross-section for the CERs on $^{100}$Mo. The ratios for the CERs to the 838 keV 1$^+$ and 1339 keV 1$^+$ states are plotted as a function of the $V^{\rm T}$/$V^{\rm C}$ in Fig. 6.  They suggets a small T contribution for the CER to the 838 keV state, and no appreciable T NME for the CER to the 1339 keV state.  

The CER on $^{128}$Te  to the $^{128}$I 1$^+$ ground state is the spin stretched SQP transition of the 2d3/2$^+$ neutron $\rightarrow$2d5/2$^+$ proton. The measured NMEs of $M$(GT)=0.295 is just the SQP ones and nearly same as the EC/$\beta$ value \cite{ele15}. Thus we take it as a reference cross-section for the CERs on $^{128}$Te. The ratios for the CERs to the strongly excited  1153 and 1222 keV 1$^+$ states in the $^{100}$Tc remain same as $V^{\rm T}$/$V^{\rm C}$ increases. Thus they are mainly the GT ones.


\begin{figure}[h]
\begin{center} 
\hspace{-0.cm}
\vspace{-0cm}
\includegraphics[width=0.33\textwidth]{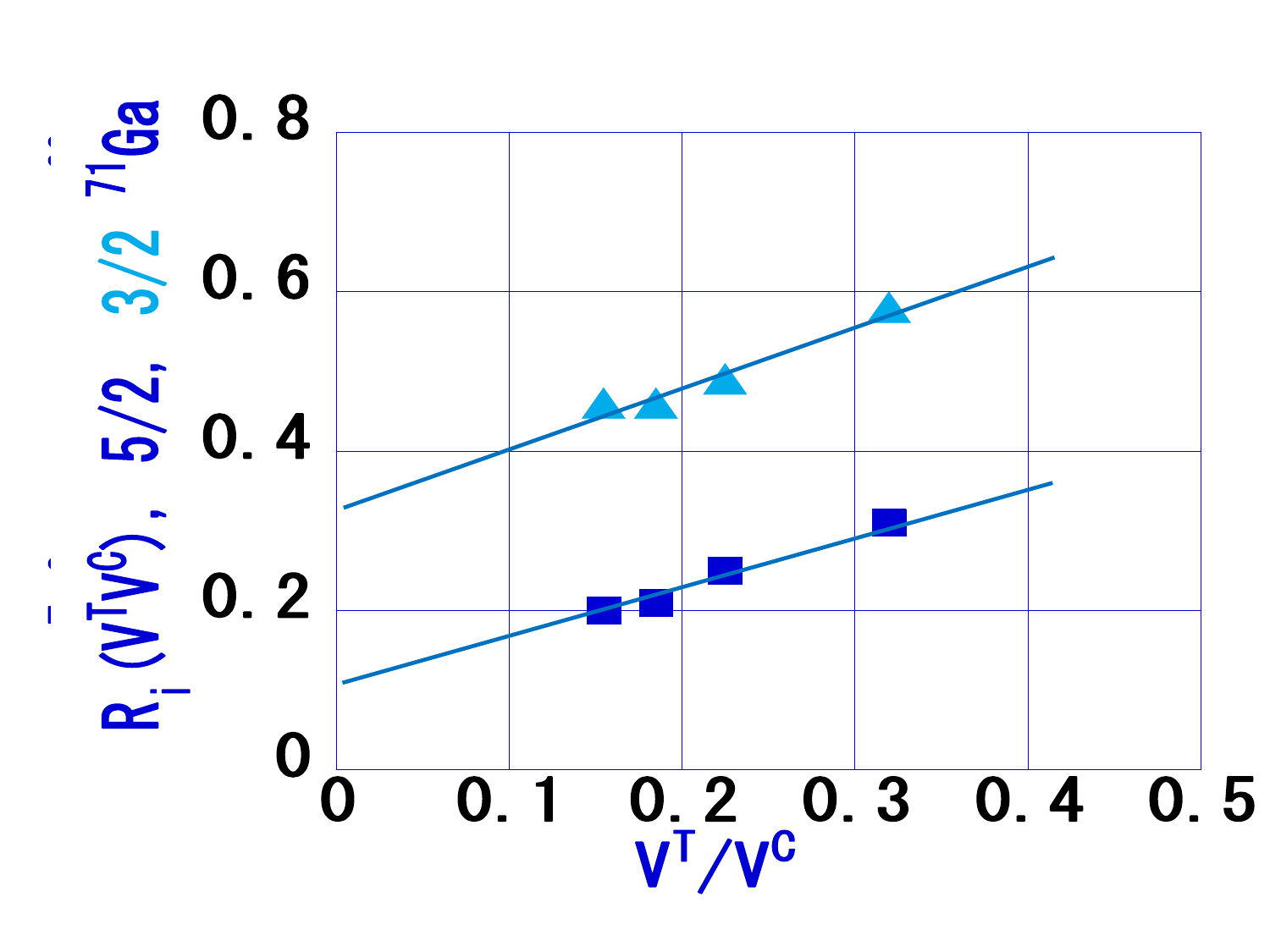}
\vspace{-.6cm}
\end{center} 
\caption{The ratios $R_i(V^T/V^C)$ for the CERs on $^{71}$Ga to the 5/2$^-$ 175 keV (squares) and 3/2$^-$ 500 keV (triangles) states in $^{71}$Ge. The errors are as given by the size for each point, and the straight lines are fits to the data. 
\label{figure:fig1}} 
\vspace{-0cm}
\end{figure}

\begin{figure}[h]
\begin{center} 
\hspace{0.3cm}
\vspace{-0.2cm}
\includegraphics[width=0.33\textwidth]{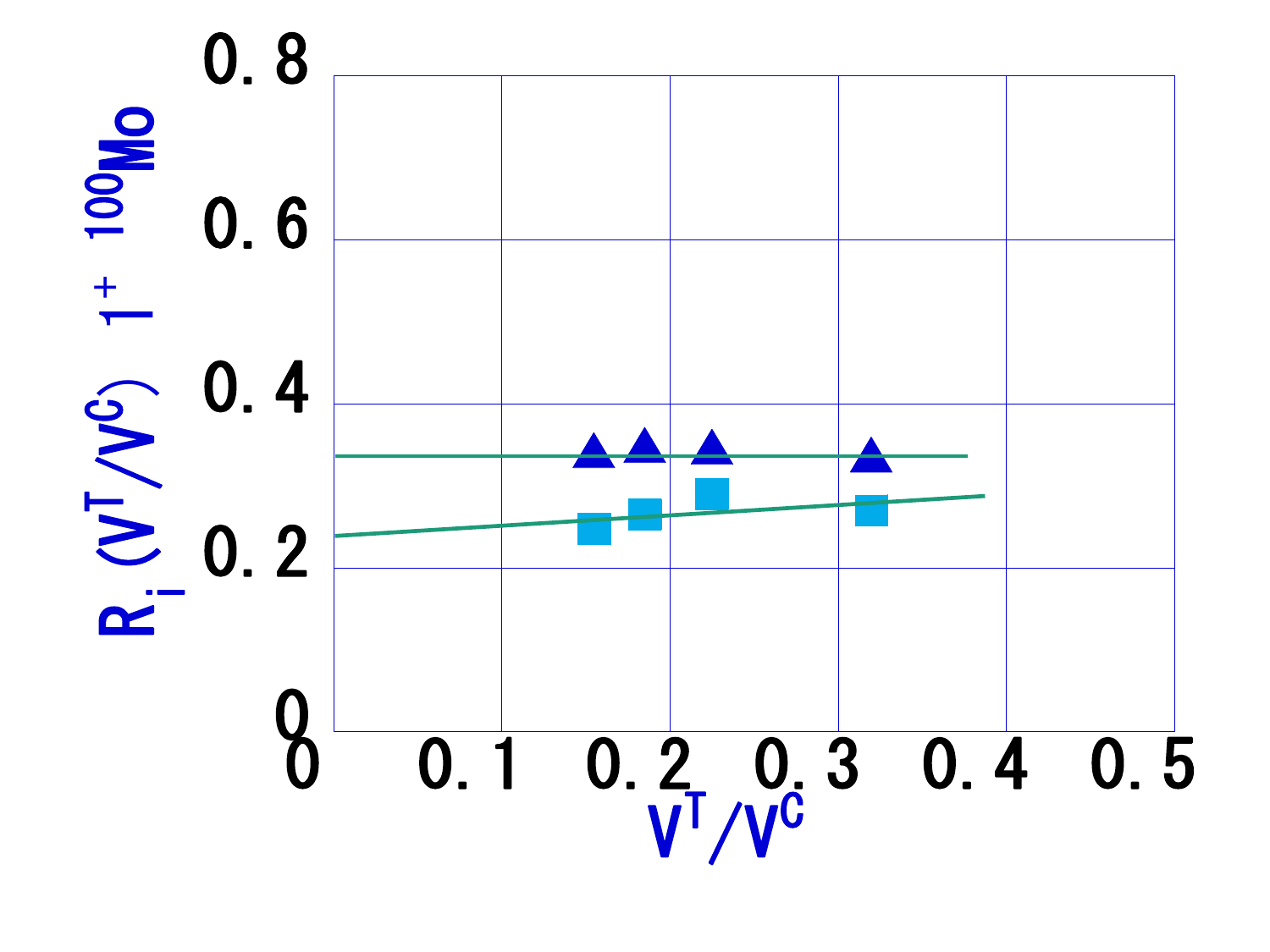}
\vspace{-0.7cm}
\end{center} 
\caption{Ratios $R_i(V^{\rm T}/V^{\rm C})$ for the CERs on $^{100}$Mo to the 1$^+$ 838 keV (squares) and 1339 keV (triangles) states in $^{100}$Tc.  The errors are as given by the size for each point. The straight lines are fits to the data. 
\label{figure:fig1}} 
\vspace{-0.8cm}
\end{figure}

\begin{figure}[h]
\begin{center} 
\hspace{1.5cm}
\vspace{-1.5cm}
\includegraphics[width=0.42\textwidth]{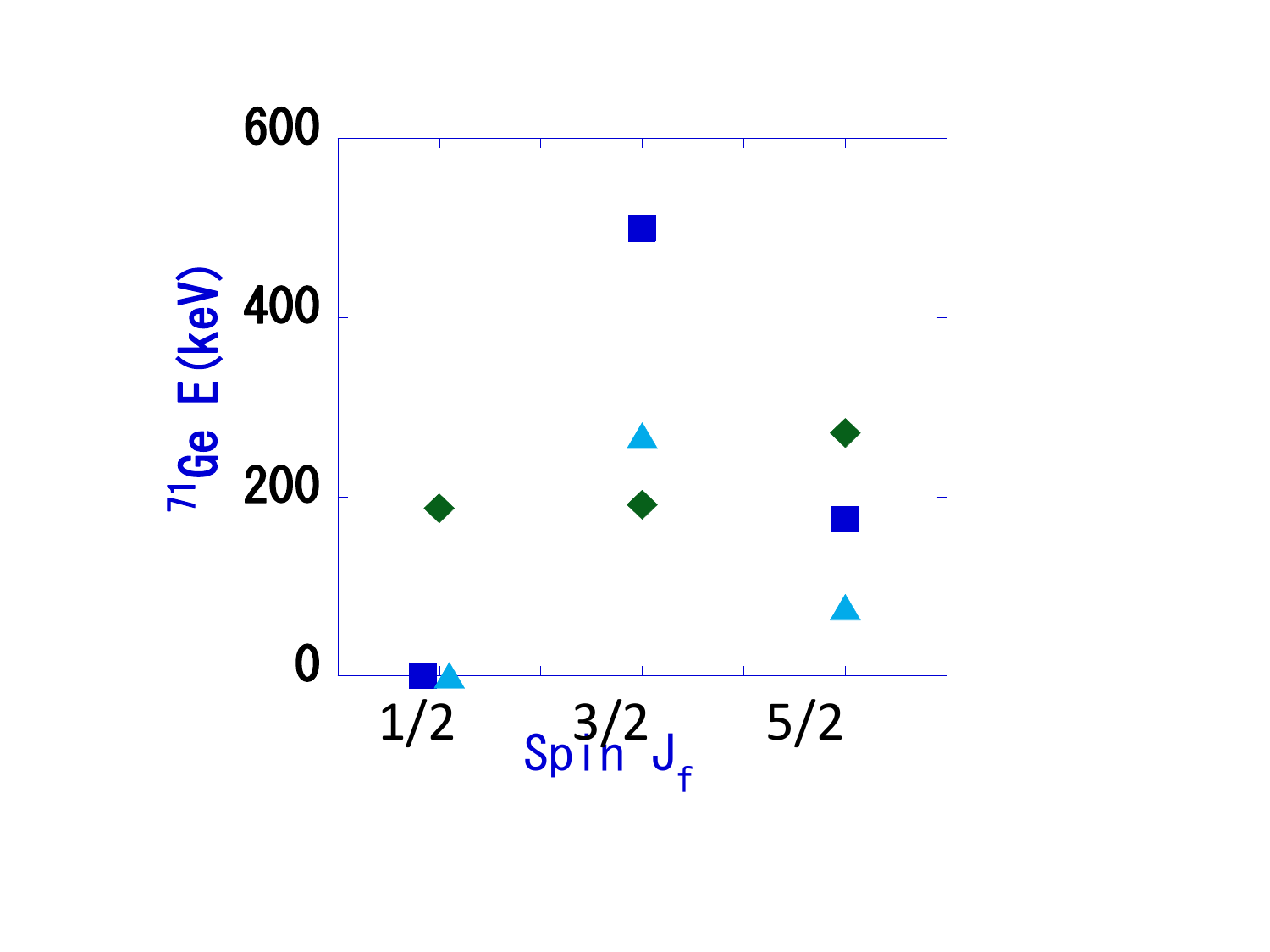}
\vspace{-0.1cm}
\end{center} 
\caption{ Energy levels of the lowest 3 states. Experimental excitation energies (blue squares), the JUN45 ISM ones (light blue triangles) and the jj44b ISM ones (green diamonds).
\label{figure:fig1}} 
\vspace{-0.3cm}
\end{figure}

\begin{figure}[h]
\begin{center} 
\hspace{.5cm}
\vspace{-1.5cm}
\includegraphics[width=0.53\textwidth]{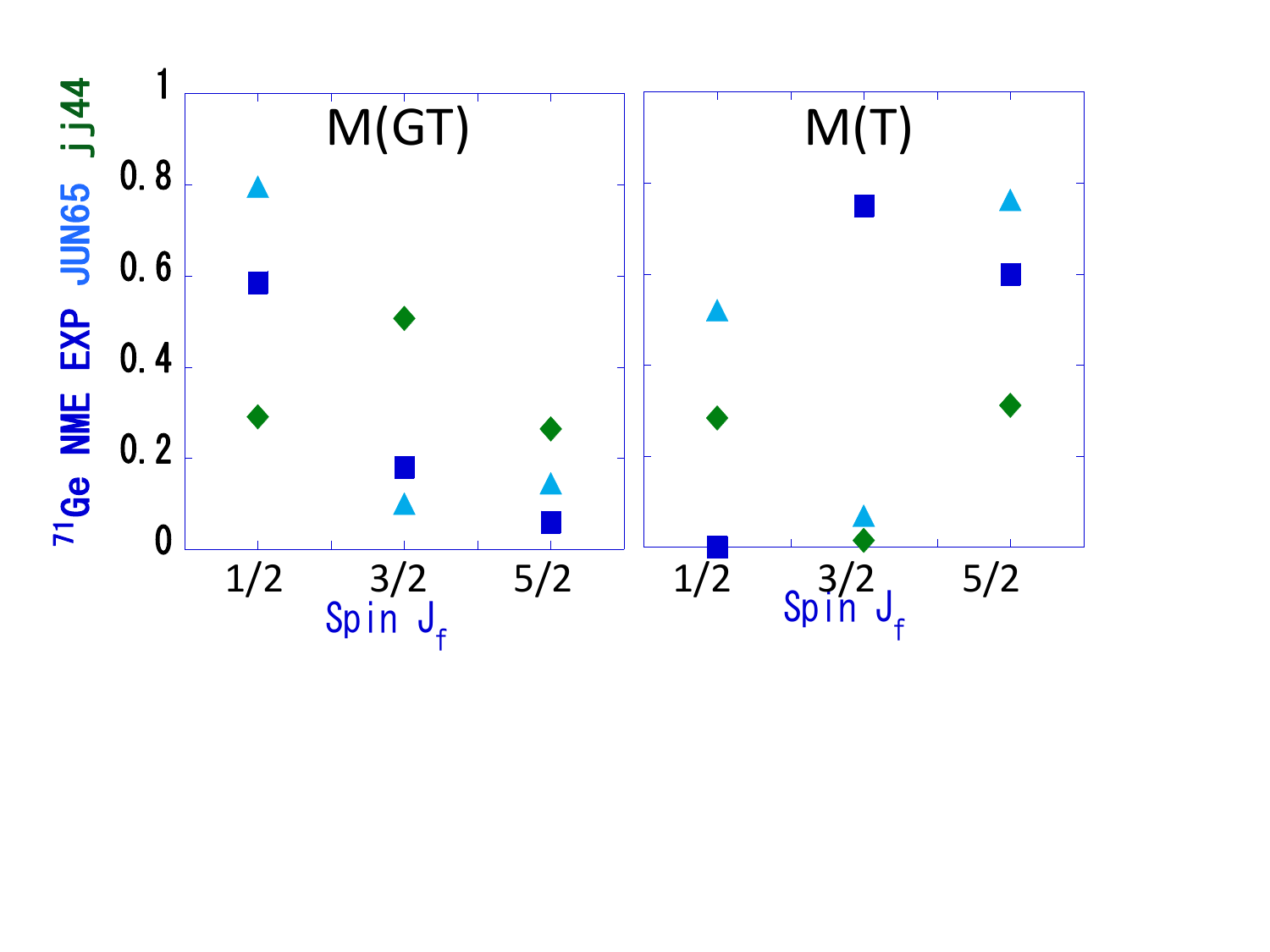}
\vspace{-1.8cm}
\end{center} 
\caption{ $M$(GT) (left pannel) and (8$\pi)^{1/2}M$(T) (right pannel)  for the lowest  3 states in $^{71}$Ge. Experimental (blue squares), the JUN45 ISM (light blue triangles) and the jj44 ISM (green diamonds).
\label{figure:fig1}} 
\vspace{-0.3cm}
\end{figure}

\begin{table}[hb]
\vspace{-0.5cm}
\caption{CERs on $^{71}$Ga, $^{100}$Mo and $^{128}$Te, and the GT and T NMEs for the low lying states in $^{71}$Ge, $^{100}$Tc and $^{128}$I. 
$E_f$(keV): excitation energy.  $J_f$: final state spin. $M_{i}$(GT): $i$ th final state GT NME. $M_{0}$(GT)=$M_{r}$(GT): $M_{i}$(T): final state T NME.
\label{tab:1}}
\vspace{0.3cm}
\hspace{-0.cm}
\begin{tabular}{ccccc}
\hline
$^A_Z$X&$E_{\rm f}$(keV) & $J_{\rm f}$&   $M_{\rm i}$(GT)  & $M_{\rm i}$(T) \\
\hline
$^{71}$Ge  &0 & 1/2$^-$  &~~0.585$\pm{0.02}$&~~~ -~~~\\
$^{71}$Ge  &175 & 5/2$^-$&~~~0.06$\pm{0.02}$&~~~0.12$\pm{0.04}$\\
$^{71}$Ge  &500& 3/2$^-$&~~~0.18$\pm{0.03}$& ~~~0.15$\pm{0.04}$\\
$^{100}$Tc &0 &      1$^+$&~~0.590$\pm{0.02}$&~~~-~~~\\
$^{100}$Tc &838&  1$^+$&~~~0.13$\pm{0.02}$& ~~~0.01$\pm$ 0.007\\
$^{100}$Tc  &1339&  1$^+$&~~~0.20$\pm{0.03}$& ~~~$\le$ 0.03\\

$^{128}$I &0       & 1$^+$&~~~0.295$\pm{0.02}$ &~~~- ~~~\\
$^{128}$I &1153  & 1$^+$&~~~0.30 $\pm{0.02}$ & ~~~$\le$ 0.03\\
$^{128}$I & 1222 & 1$^+$&~~~0.35$\pm{0.02}$  & ~~~$\le$ 0.03\\

\hline
\vspace{-0.3cm}
\end{tabular}
\end{table}

The present analyses indicate that the postulated ratios at $V^{\rm T}$/$V^{\rm C}$=0 are smaller than the observed ones at $\theta \approx q$=0 (i.e. $V^{\rm T}$/$V^{\rm C}$=0.14) in cases of the 175 keV 5/2$^-$ and 500 keV 3/2$^-$ states in $^{71}$Ge, as shown in Fig.5, but they are nearly equal to the observed ones there in cases of the $^{100}$Tc and $^{128}$I as shown in Fig. 6 and Fig.7.   Accordingly, the $M$(GT)s derived from the ratio at $V^{\rm T}$/$V^{\rm C}$=0, i.e. the GT NMEs corrected for the tensor contributions, are smaller than the $M$(GT)$(1+f(q))$
 derived from the ratio at $V^{\rm T}$/$V^{\rm C}$=0.14 (namely at $q\approx \theta$=0) in cases of the $^{71}$Ge. They are nearly same as $M$(GT) at $V^{\rm T}$/$V^{\rm C}$=0.14 ($\theta$=0 deg.) in cases of the two states in $^{100}$Tc  and $^{128}$I.

The tensor NME of $M$(T)s is derived experimentally from  the observed gradient of $\epsilon M_i({\rm T})$/$M_i({\rm GT})$ with $\epsilon V^{\rm T}$/$V^{\rm C}$=0.42 at $q$=0 (0 deg.). This is in accord with the tensor coefficients around 0.26 in \cite{iwa26}, 0.38 in \cite{ell24} and 0.5 in \cite{kos19}.
The obtained NMEs of $M^{\rm T}$ are shown in Table 1. 

Here we make a few remarks on the NMEs from the CERs.
i) The $M$(GT) and $M$(T) derived experimentally reflect the shell configurations of the states. The $M$(GT) for the 5/2 state in $^{71}$Ge is as small as 0.06, and the $M$(T) is comparable with the $M$(GT). They are explained by the $l$-forbidden 1f5/2 $\rightarrow$ 2p3/2 
 component, which favors the $M$(T) rather than $M$(GT).  
ii) Interesting is to note that the observed small $M$(GT)$\approx$0.11$\pm0.03$ for the ground state of Ge$^{69}$ agrees with the $M$(GT)$\approx$0.09 derived from the EC experiment. 
iii) The CERs on $^{100}$Mo to the low-lying states in $^{100}$Tc and those in $^{128}$I are mainly SQP transitions of (1g7/2)$_n\rightarrow$(1g9/2)$_p$ and those of (1d3/2)$_n\rightarrow$(1d5/2)$_p$, and
 no $l$-forbidden transition is involved there. Then the CERs to the strongly excited 1339 keV state in  $^{100}$Tc and 1153 keV and 1222 keV states in  $^{128}$I are mainly GT transitions, but the weakly excited 838 keV in  $^{100}$Tc may include a little T contribution.

Now, let us compare the present results with the shell model calculation. The energies and the GT and T NMEs for the low states in $^{71}$Ge were calculated by using the JUN45 and jj44b ISM (Interacting Shell Model) codes \cite{hom09,che10}. The excitation energy of the energy level and the GT/T NMEs reflect the shell configuration of the state.  The  calculated excitation energies for the lowest 3 states with spin 1/2, 3/2, and 5/2 are compared with the experimental ones in Fig.7.  Important is to note that the calculated excitation energies by the JUN45 and jj44b ISMs are very different from each other, and are also very different from the experimental ones. 

Accordingly, the GT and T NMEs  calculated by the  JUN45 and jj44b ISMs are very different from each other, and they are also very different from the present experimental values, as shown in Fig.8. One critical point of the ISM is that it does not explicitly include  the $\Delta$-resonance (non-nucleonic) effect, which reduces the GT and T NMRs by a factor around $g_{\tau\sigma}^{eff}$/$g_{\tau\sigma}\approx$ 0.7 with $g_{\tau\sigma}$ being the isospin spin (axial vector) coupling \cite{eji25}. Using this coefficient, the JUN45 NME for the (1/2)$^-$ ground state in $^{71}$Ge agrees with the experimental one, but all others are not. \\ 

{\it Concluding remarks and discussions. }~The medium energy CER is a realistic probe to study experimentally GT, SD and other NMEs associated with DBD and IBD $\nu$ studies \cite{eji00,eji19}. The ($^3$He,$t$) CER with the 0.42 GeV $^3$He projectile at RCNP has been extensively used to study   them. The CER with $\Delta l$=0 via the nuclear interaction, however, includes both the GT excitation by the central interaction $V^{\rm C}$ and the T excitation by the tensor interaction $V^{\rm T}$. They interefere with each other. 

So far  $B(\rm GT)$ ($M$(GT)$^2/(2J_{\rm i}+1)$) has been derived experimentally  from the CER cross section $\sigma$(0) at $\theta \approx$0 deg. (i.e. $q\approx$0), assuming no interference with the tensor term, relying on the so-called the proportionality relation, i.e. $\sigma(0) =k B$(GT) \cite{eji00,goo80,tad81}. The relative uncertainty due to the possible tensor contribution is shown to be around 0.03-0.035 ln($B(\rm GT)$), i.e. 10$\%$ for $B(\rm GT)\approx$0.1 (i.e. $M$(GT)$\approx$0.3), but amounts to 20$\%$ for $B(\rm GT)$=0.01 (i.e. $M$(GT)$\approx$0.1) \cite{zeg06,hit09}. Thus the CER  has been considered to be appropriate only in limited cases of $B(\rm GT)\ge$0.1, namely $M$(GT)$\ge$0.3, as the refernce GT NMEs used in the present analysis.

The present work shows for the first time how both the $M$(GT) and $M$(T) in the wide excitation region are derived experimentally from the CER cross sections in the forward region ($\theta$=0-2 deg.). The differential cross section for the CER with the relative $\Delta l$=0 (i.e. s-wave $t$) decreases as $|j_0(qR)V^{\rm C}(1+\delta V^{\rm T}/V^{\rm C})|^2$ with $\delta=\epsilon M$(T)/$M$(GT). Noting that $V^{\rm T}$/$V^{\rm C}$ increases rapidly as $q$ increases,  the tensor contribution manifests itself in the $q$ distribution of the differential cross section. Actually, the $q$ distribution is a little modified by the distortion potential, and thus the analysis is made by using the DWBA (Distorted Wave Born Approximation).
The $M$(GT) and $M$(T) are derived by comparing the measured $q$ (i.e. $\theta$) distribution with the measured  reference one for the spin-stretched SQP state excited mainly by the GT excitation. The used $M_{\rm r}$(GT) agrees with the GT NME measured by the EC/$\beta$ within 5 $\%$. 
 In fact, there is a small (2-5$\%$) T contribution even in the spin-stretched SQP NME, but the effect is much smaller than the unsertainties in the $M_i$(GT) in Table 1 and typical experimental uncertainty in the $M$(GT).

The CER angular distribution is modified more or less by the nuclear distortion potential. Then, the present method of using a SQP NMR as a reference GT one can be applied for states nearby the SQP in the excitation energy and the mass number. Otherwise one needs accurate analyses of angular ($q$) distributions in terms of the DWBA. Here we note that the experimental cross-sections at the forward angles of 1-2 deg. include the small ( around 5 $\%$) contribution from the $\Delta L$=2, which is corrected for by using the DWBA fit. 

In the previous work, we have shown that the $M_i$(GT)s for the lowest 3 states in $^{71}$Ge are obtained from the CER $M$(110)s corrected for the T contributions estimated by the ISM calculation \cite{iwa26}. They are  barely within the experimental errors of the present values for the $M$(GT)s. The present way is based on experiments, being free from unknown uncertainty involved in the theoretical calculation.

The contribution of the 1st and 2nd excited states in $^{71}$Ge to the $^{51}$Cr-$\nu$ and the solar-$\nu$ cross-sections are evaluated by using the present experimental B(GT) values. The ratio to the contribution from the ground state is $K$=1.028$\pm$0.003. This is consistent with the previous values \cite{ell23,iwa26}, and shows that the Ga anomaly\cite{abd06,bar22a,bar22b} is not due to the nuclear response. This would suggest oscillation to a sterile neutrino on the basis of the experimental observations \cite{bar22a,bar22b}.

Acturelly,  GT and T NMEs reflect the nuclear configurations involved in the transition, namely spin streched and spin non-stretched SQP transitions, quadrupole phonon transitions, $l$-forbidden transitions and so on. Then the GT and T NMRs studied by CERs are used to investigate the nuclear configurations and nuclear structures involved in their NMEs.  

It is remark that  GT and T NMEs evaluated theoretically by using model wave functions are very sensitive to nuclear models and nuclear parameters used for the NME calculations \cite{eji19,iwa26}. The JUN45 and jj44b ISM calculations for Ga-Ge isotopes do not reproduce the experimental energies and GT/T NMEs. GT and T NMEs derived experimentally, together with  realistic model calculations, are valuable to investigate the nuclear structures involved in the CERs.

\end{document}